\documentclass[10pt,conference,compsocconf]{IEEEtran}
\IEEEoverridecommandlockouts
\usepackage{amsmath,amssymb,amsfonts}
\usepackage{algorithmic}
\usepackage{graphicx}
\usepackage{textcomp}
\usepackage{xcolor}
\usepackage{cite}
\usepackage{makecell}
\usepackage{svg}
\usepackage{breakurl}
\PassOptionsToPackage{hyphens}{url}\usepackage{hyperref}

\hypersetup{
    colorlinks=true,
    linkcolor=magenta,
    filecolor=magenta,
    urlcolor=magenta,
    citecolor=magenta,
}

\usepackage{subcaption}
\usepackage{threeparttable}
\usepackage{mathtools}
\usepackage[labelfont={bf,small},textfont={small}]{caption}
\usepackage{xspace}
\usepackage{balance}
\usepackage{booktabs}

\def\BibTeX{{\rm B\kern-.05em{\sc i\kern-.025em b}\kern-.08em
    T\kern-.1667em\lower.7ex\hbox{E}\kern-.125emX}}

\makeatletter
\def\ps@IEEEtitlepagestyle{%
  \def\@oddfoot{\mycopyrightnotice}%
  \def\@oddhead{\hbox{}\@IEEEheaderstyle\leftmark\hfil\thepage}\relax
  \def\@evenhead{\@IEEEheaderstyle\thepage\hfil\leftmark\hbox{}}\relax
  \def\@evenfoot{}%
}
\def\mycopyrightnotice{%
  \begin{minipage}{\textwidth}
  \scriptsize
  \copyright~2021 IEEE Personal use of this material is permitted. Permission from IEEE must be  obtained for all other uses, in any current or future media, including  reprinting/republishing this material for advertising or promotional purposes, creating new  collective works, for resale or redistribution to servers or lists, or reuse of any copyrighted  component of this work in other works.
  \end{minipage}
}
\makeatother

\newcommand{\para }[1]{\medskip \noindent  {\bf #1}}
\newcommand{\1}{{\em (i)}}
\newcommand{\2}{{\em (ii)}}

\newcommand{\sysname}{\textsc{DELI}\xspace} %

\newif\ifhiresimg
\hiresimgtrue
\newcommand{\altimages}[2]{\ifhiresimg#1\else#2\fi}

\begin{document}

\newcommand{\issue}[1]{{\color{red} (#1)}}
\newcommand{\tian}[1]{{\color{purple}\textbf{Tian:} \textit{#1}}}
\newcommand{\note}[1]{}
\newcommand{\stkout}[1]{\ifmmode\text{\sout{\ensuremath{#1}}}\else\sout{#1}\fi}
\makeatletter
\newcommand{\thickhline}{%
    \noalign {\ifnum 0=`}\fi \hrule height 1pt
    \futurelet \reserved@a \@xhline
}
\makeatother

\title{Quantifying and Improving Performance of Distributed Deep Learning with Cloud Storage}

\author{
\IEEEauthorblockN{Nicholas Krichevsky\IEEEauthorrefmark{1}, Renee St Louis\IEEEauthorrefmark{2}, Tian Guo\IEEEauthorrefmark{3}}
\IEEEauthorblockA{\textit{Computer Science Department} \\
\textit{Worcester Polytechnic Institute}\\
Worcester, MA, USA \\
\IEEEauthorrefmark{1}njkrichevsky@wpi.edu, \IEEEauthorrefmark{2}rstlouis@wpi.edu, \IEEEauthorrefmark{3}tian@wpi.edu
}
}
\maketitle

\begin{abstract}
Cloud computing provides a powerful yet low-cost environment for distributed deep learning workloads. However, training complex deep learning models often requires accessing large amounts of data, which can easily exceed the capacity of local disks. Prior research often overlooks this training data problem by implicitly assuming that data is available locally or via low latency network-based data storage. Such implicit assumptions often do not hold in a cloud-based training environment, where deep learning practitioners create and tear down dedicated GPU clusters on demand, or do not have the luxury of local storage, such as in serverless workloads. In this work, we investigate the performance of distributed training that leverages training data residing entirely inside cloud storage buckets. These buckets promise low storage costs, but come with inherent bandwidth limitations that make them seem unsuitable for an efficient training solution. To account for these bandwidth limitations, we propose the use of two classical techniques, namely caching and pre-fetching, to mitigate the training performance degradation. We implement a prototype, \sysname, based on the popular deep learning framework PyTorch by building on its data loading abstractions.  We then evaluate the training performance of two deep learning workloads using Google Cloud's NVIDIA K80 GPU servers and show that we can reduce the time that the training loop is waiting for data by 85.6\%--93.5\% compared to loading directly from a storage bucket---thus achieving comparable performance to loading data directly from disk---while only storing a fraction of the data locally at a time. In addition, \sysname has the potential of lowering the cost of running a training workload,  especially on models with long per-epoch training times.

\end{abstract}


\section{Introduction}
    Cloud computing is a promising low-cost frontier for distributed deep learning. However, as deep learning practitioners explore more complicated problems that require increasingly large datasets, the choice of where to store this training data can drastically affect both training time and training cost. For example, when using SGD-based algorithms to train a deep learning model, each GPU worker in the cluster needs access to a subset of the training data. Prior research often overlooks the problem of where to store the large amount of training data by implicitly assuming that data is available locally or via low latency network-based data storage~\cite{bigdl_dai,pytorch_distributed_li}. Such implicit assumptions often do not hold in a cloud-based training environment, where deep learning practitioners create and tear down dedicated GPU clusters on demand. Furthermore, in emerging training scenarios such as serverless-based training~\cite{Cirrus} (where workers only have access to ephemeral storage) or online learning~\cite{elias:wheres-the-bear} (where training data is collected in real-time), centralized storage such as cloud bucket storage remains the only option.
    However, leveraging cloud bucket storage requires additional mechanisms to handle its inherent bandwidth constraints and to ensure a minimal performance penalty compared to training from disk.

    Distributed deep learning requires that each node be able to access the training data, which can be accomplished through storing it locally ahead of time or accessing it from a remote store. In the latter case, workers may end up waiting a non-negligible amount of time for samples to download before they can continue training. While storing all training data locally ahead of time may seem an obvious solution, this is not always feasible. Storing the entire dataset on each node may be prohibitively expensive; the cost of doing so is proportional to the number of nodes it is stored on, while bucket storage requires only paying for storage once. Additionally, doing so may not be compatible with certain architectures. Serverless environments, for instance, may necessitate using bucket storage because they have limited, ephemeral storage, rather than large amounts of persistent storage \cite{Cirrus,website:lambda_storage, wang:serverless-SIREN, jonas:pyWren-cloud-communism, kurz:double-machine-learning}. Additionally, online learning can require large amounts of real-time data, making storage buckets an apt choice for cost-effectiveness and upload speed \cite{elias:wheres-the-bear, alipour:microservice-logs, makkie:mri, ma:medhere}.

    In cases like these, bucket-based object storage is advantageous for providing a globally accessible location for storing large quantities of data, but loading these samples during training can become a bottleneck that slows the process. A system that can address this problem could reduce data loading time, which in turn would maximize the usefulness of cloud resources. This could make serverless and online learning architectures more viable, and it would have the potential to cut costs in cloud-based distributed deep learning.

    In this work, we set out to \emph{investigate the feasibility of and improve the performance of loading training data from storage buckets for distributed deep learning}. In general, certain data storage strategies dramatically increase data loading time, and ergo, total training time. As such, our research concentrates on the \emph{data loading process}: the pipeline of loading training data from a storage medium and preparing it for training.

    We begin our analysis of this problem by showing why it is not an acceptable solution to naïvely read data from a storage bucket when the training worker needs it. Our empirical measurement demonstrates that the training performance degrades significantly when using cloud buckets. As such, we design a system called \sysname that leverages two classical approaches---caching and pre-fetching---to improve the data loading time in a cloud-based distributed training setting. We evaluate the distributed training performance using \sysname on top of Google Cloud Platform, showing significant data loading speedup and potential cost saving. To summarize, we make the following main contributions:
    \begin{itemize}
        \item We design a configurable caching and pre-fetching system, called \sysname, for loading training data from cloud bucket-based storage. \sysname requires only a fraction of the training data to be stored on the local disk at a time. \sysname's source code and our raw experimental data can be found at  \url{https://github.com/cake-lab/DELI}.
        \item We find an extremely effective configuration of our system \sysname, which we experimentally observed to speed up data loading by 85.6\% to 93.5\% compared to loading from a storage bucket. This configuration works by pre-fetching training data as soon as half of the cache has been trained on.
        \item We develop a simple cost model based on Google Cloud Platform's pricing scheme and our empirical measurements, and we show that \sysname could potentially save money in certain situations.
    \end{itemize}

    While many researchers have explored optimizing coordination (e.g.\ gradient synchronization~\cite{zhao2019dynamic,li2021syncswitch_icdcs}) little work has been done to measure and enhance data loading, and what work has been done usually does not focus on cloud environments \cite{bigdl_dai,pytorch_distributed_li}. That said, Cirrus, a serverless framework for machine learning, is a closely related work to ours. Cirrus' focus, however, is on overall training performance, rather than data loading time, which is this paper's focus~\cite{Cirrus}.

\section{Background}
\label{chapter:background}

    \subsection{Distributed Deep Learning}

    Distributed deep learning amplifies the power of machine learning by using multiple nodes to train complicated models. Just like its single-node counterpart, training is completed by feeding samples of data through a neural network and computing gradients of model output. The parameters of the model are then updated using these gradients. To improve efficiency, the set of samples is often broken into \emph{mini-batches} (small, disjoint subsets of the dataset), and training over each mini-batch once constitutes an \emph{epoch}. Distributed deep learning builds on top of this by allowing training to be split across multiple nodes. This can allow for \emph{model parallel} training, wherein each node contains only a subset of the whole model, and \emph{data parallel} training, wherein each node trains on a subset of the data rather than the whole dataset, should either be useful.

    While many possible architectures may suit the needs of this research, we have chosen to use a distributed data parallel architecture, wherein each node uses the full model but only a subset of the data. To synchronize the model between workers, each worker periodically communicates with one worker that is designated as the \emph{master}. This master worker will accumulate all of the gradients, and then communicate them back to the other workers, using a process known as AllReduce \cite{pytorch_distributed_li}. Specifically, we used PyTorch's \texttt{DistributedDataParallel} package due to its battle-tested implementation. We rely on this package's resilience on gradient communication so that we may focus our energy on data loading. Ultimately, our choice of architecture should not affect our findings noticeably because the system we implement is agnostic of its surrounding architecture. A parameter server architecture, wherein nodes periodically synchronize parameters with a centralized server rather than in a ring, should work just as well.

    In the distributed data parallel architecture we have chosen to study, there are two components of communication: gradient synchronization and data loading. In short, the gradient synchronization step is used to ensure that the model is updated using the calculations made by each worker, maintaining mathematical accuracy. The data loading step ensures that each worker has some kind of data to train on \cite{pytorch_distributed_li}.

    For each step of the training process, each node will receive a single mini-batch to train on. While this data may be loaded onto each node ahead of time if the resource budget and architecture allow it, this is not always possible. If the dataset is exceedingly large, or if the system is constrained by other technical factors, this data has to be stored elsewhere, and mini-batches must be downloaded by each node during training. Downloading this data takes time, and if the transfer of training data for the next step is not completed before the current step's computation completes, the next step's computation will be forced to wait.

    \subsection{Problems with Data Loading}
        \label{section:moving-to-cloud}

        When training a large model in a distributed fashion, it is not uncommon to have access to a high performance computing (HPC) cluster. These clusters can be configured such that nodes have uninterrupted high-bandwidth connections with each other \cite{scaling_dl_naumov}. This level of control allows those who use HPC clusters to efficiently retrieve their training data from a single networked file system (e.g.\ NFS) \cite{data_transfer_yang}. Unfortunately, configuring such a cluster requires a significant up-front investment, making a cloud-based solution much more preferable.

        In general, reading training data directly from disk is quite fast---on the order of 100 MB/s for a spinning magnetic drive. However, using them in the cloud requires paying for storage on each node. Using cloud object storage sidesteps this problem, but performance is orders of magnitude lower, and sometimes unpredictable. To make matters worse, cloud providers, such as Google Cloud Platform (GCP), do not allow the ability to download multiple files at once, but rather one at a time~\cite{website:gcp-batching}, which further hurts performance. To demonstrate this, we use one of the VMs described in Section \ref{chapter:results} to read each image from the MNIST dataset into memory. We tested two setups: \1 reading sequentially from both object storage and disk, and \2 reading with 16 threads performing fetches from object storage.  The averaged results are shown in Table \ref{table:mnist-read}. Note that the disk speed is a fraction of the previously mentioned speed; reading many small files carries more overhead than a sustained sequential read.

        \begin{table}[t]
            \caption{Speed of reading the MNIST dataset into memory, both sequentially and in parallel.}
            \label{table:mnist-read}
            \centering
            \begin{tabular}{l  r  r}
                \toprule
                \textbf{Data Source} & \textbf{Transfer Speed} & \textbf{Std. Dev.} \\
                \hline
                Disk & 18.63 MB/s & 0.19 MB/s \\
                Object Storage (Sequential) & 49.80 kB/s & 3.85 kB/s \\
                Object Storage (Parallel) & 281.73 kB/s & 4.29 kB/s \\
                \bottomrule
            \end{tabular}
        \end{table}

        The difference in speeds between object storage and disk is realized when one attempts to train a model with a bucket as the training data store, as shown later in Figure \ref{fig:graph_baselines}. More details can be found in Section \ref{section:experiment-setup} but briefly, by measuring the time the training worker spends loading training data, we notice a dramatic increase in loading time.

\section{Design and Considerations}
\label{chapter:approaches}

In general, we want to find ways to lower the amount of time that cloud-based workers spend waiting to load training data from a storage bucket. In this section, we describe our design considerations for achieving this in \sysname, which employs two classical techniques: caching and pre-fetching of training data. We also develop a cost model to better understand the monetary implications of using \sysname.

\subsection{Caching}
\label{section:cache-design}

As discussed in Section \ref{section:moving-to-cloud}, the time needed to load data from object storage significantly increases the training time of the model. The most obvious way to reduce this time is by storing as much of the data locally as we can---thereby reducing time-consuming network traffic. As such, we add a caching layer to each training worker, which will store each sample after the worker requests it. In its simplest form, this can store samples as they are trained on, or samples can be cached ahead of time.

In designing our caching approach, we decided on three main constraints. First, even though networked caches are traditionally stored in-memory, memory is far more expensive than storage in cloud environments \cite{gcp_compute_storage_pricing, gcp_compute_memory_pricing}. As such, we wanted to make sure that our database was disk-based. Second, to minimize the total number of network requests, thus maximizing the speed advantage of caching, we decided that the caches would be placed on each node. This way, when a node wants to retrieve cached data, it would only need to wait for the database and not for the data to travel across the network. Last, we wanted to keep our API simple-to-use and non-invasive, so we decided to encapsulate as much of our implementation within PyTorch's abstractions as possible.

\subsection{Pre-fetching}
\label{section:prefetch-design}

Our goal in pre-fetching is to minimize network traffic and to overlap communication and computation as much as possible. With this in mind, we chose to let each node manage its own local cache independently. Each node independently determines which samples it wants, then requests them from the storage bucket while training on other data.

 To implement pre-fetching, we created a service that would fetch the samples from the bucket ahead of when they would be needed for the training loop, and load them into the cache. Each node hosts its own instance of this service, removing concerns of network delays. The training loop will periodically inform this service of which samples it plans to train on. The number of samples that the training loop requests is configurable, hereby referred to as the \textit{fetch size}. In typical usage, the fetch size is only a fraction of the actual dataset size, but as Section~\ref{section:prefetch-results} demonstrates, the larger the value of the fetch size, the better. Upon receiving this request, the pre-fetch service then caches these samples asynchronously while training continues. The training loop has no knowledge of the fetch's completion; it will simply attempt to retrieve the samples from the cache if possible, or fall back to the bucket if needed.

 In the default configuration, the training loop will train on all fetched samples before requesting another fetch from the pre-fetch service. In order to ensure more data is ready ahead of time, we introduce a new parameter: the \textit{pre-fetch threshold}. This parameter adds a buffer of sorts---a minimum number of samples that have been fetched but not trained on. Once the training loop crosses this threshold, another fetch is made, earning this number the name of pre-fetch threshold.

\subsection{Cost Analysis}
\label{section:cost-design}
To ensure our system would be cost-effective, we analyze the cost of using buckets as a storage medium for distributed training. First, we must consider the components of running the training VMs that incur a cost: storage space \( s \), number of nodes \( n \),  and runtime \( t \). Storage can be broken down into two main components: the space to store the operating system and all necessary dependencies \( s_r \), and the space needed to store the training data (i.e.\ the size of the dataset) \( s_t \). Similarly, runtime can be broken down into the time needed to perform computations \( t_c \), and the time needed for training data to load \( t_d \).\footnote{It is assumed that both of these are non-overlapping. In other words, \( t_d \) is the amount of time that the training loop spends waiting to perform further computations.} The amount of time to load data can vary depending on available bandwidth, dataset used, or model used\footnote{The model is a factor here because the overlapping computation and communication time mean that given a long-running model, the training loop may be waiting for data for less time.}; therefore, we recommended that \(t_d\) be measured empirically, either by performing a full training run and measuring the time spent loading data, or by loading a couple of batches of data, and multiplying out how long it would take to load all batches of data. However, \( t_c \) can be estimated theoretically using a model to estimate training time \cite{lee:modelling_time1, shi:modelling_time2, oyama:modelling_time3, Yan:modelling_time4, zhou:modelling_time5}.

Second, we must take into consideration how our cloud platform of choice, GCP, bills these charges. Google charges a fixed hourly rate \(c_c\) for running each virtual machine, and a fixed monthly rate \(c_d\) for storage on each virtual machine. We can use this information to express the expected cost of running a training workload as follows:

\begin{align}
    &n \times (c_d \times (s_t + s_r) + \tau) \\
    \shortintertext{where}
    &\tau = c_{c} \times (t_c + t_d)
\end{align}

To expand this approach to incorporate cloud storage buckets, we must consider two factors: the cost of storing the data in the bucket \( c_b \) and the cost of retrieving data from the bucket. Though within a region, data transfer is free, Google charges a separate rate for API requests, based on the type of request: Class A (which includes listing the bucket), and Class B (which includes fetching data from the bucket). Specifically, Class A requests cost \$0.05 per 10,000 requests and Class B requests \$0.002 per 10,000 requests \cite{gcp_bucket_pricing}, but we will symbolically refer to the per-request respective rates as \(c_A\) and \(c_B\). In order to determine the number of Class A requests, we must consider the number of samples, \( m \), and the number of samples returned by each listing request. We call this the page size, \( p \). The number of Class B requests, however, only depends on the number of samples, \( m \). Both of these must be multiplied by the number of epochs, \( e \), as they are repeated each epoch.

In addition, we must consider the number of samples in each node's local cache, \( m_c \). For the baseline that pulls from the storage bucket, this quantity is zero, as there is no cache to take up space.

Bringing this together yields this expression:

\begin{align}
    c_b \times s_{t} &  +\ n \times \left(c_d \times (s_r + \frac{s_t}{m} \times m_c) + \tau \right) \\
    & + 10^{-4}e\alpha \nonumber \\
    \shortintertext{where}
    \alpha & = n \times \lceil\frac{m}{p}\rceil \times c_{A}+m \times c_{B}&
\end{align}

Last, we wish to account for the pre-fetching component of our system, \sysname. To do this, we must multiply the number of listing requests by the number of fetches, as each call to the pre-fetcher requires listing the entire bucket.\footnote{The prototype we ran experiments on naïvely does this listing on every request. This is not necessarily required. In Section \ref{chapter:discussion} we discuss a possible future modification to our system that could further reduce costs by reducing the number of listing to one per node rather than one per fetch.}  The factor by which our requests are multiplied is nothing more than the ceiling of the ratio between the number of samples, \( m \) and the fetch size, \( f \). To account for this, we must simply change \( \alpha \) to the following:

\begin{align}
 \alpha = n \times \lceil\frac{m}{p}\rceil \times \lceil\frac{m}{f}\rceil \times c_{A} + m \times c_{B}
\end{align}

\section{Implementation}
\label{chapter:implementation}

We implement a prototype of \sysname, in accordance with the designs described in the previous section. This prototype is built on top of PyTorch, which was chosen due to its modularity and simplicity. Further, \sysname is currently integrated with the popular Google Cloud Platform environment for training~\cite{website:gcp_overview} but can be extended to support other cloud platforms such as Amazon Web Services. \sysname runs workers on GCP's virtual machines and holds their samples in cost-efficient storage buckets~\cite{gcp_bucket_pricing, gcp_compute_storage_pricing}.

\subsection{PyTorch's Mechanisms for Data Loading}
    \label{section:pytorch-loading}

    In order to find a place for our improvements in the data flow, we must first dissect PyTorch's data loading mechanisms. PyTorch's flow is highly modular, which gives us ample space to make our improvements. The most important pieces of this design that we can manipulate are the \texttt{Dataset}, the \texttt{Sampler}, and the \texttt{DataLoader}. The \texttt{Dataset} indexes samples and manages their lookup. The \texttt{Sampler} is a generator that selects samples strategically (e.g. sequentially, randomly, or based on node partitioning) and returns their indices for use in the \texttt{Dataset}. Finally, the \texttt{DataLoader} combines the \texttt{Sampler} and \texttt{Dataset} and, when iterated over, produces mini-batches to train on \cite{website:PyTorch_data}.

    Our implementation of both caching and pre-fetching are implemented using these primitives, and serve as drop-in replacements for their PyTorch counterparts. As the following sections describe, our implementations serve as wrappers for the user's existing \texttt{Dataset} and \texttt{Sampler}.

\subsection{Caching}
\label{section:implementation-caching}
As mentioned in Section \ref{section:cache-design}, our three primary considerations when designing the cache were that it must be on-disk, and must be available on each node. As such, we selected MongoDB as our cache-store due to its ease of use, its built-in eviction abilities, and its disk-based nature. The cache is structured as a size-limited collection (a ``capped collection'' in MongoDB parlance \cite{mongodb:capped-collections}) of entries, holding the sample itself, its index in the dataset, and a unique ID for the current training session. These two identifiers are placed in a multi-key index to speed up lookup times.

To ensure we met our goal of producing a simple API, we wanted to integrate with PyTorch's existing data pipeline (see Section \ref{section:pytorch-loading}). With this in mind, we designed a custom \texttt{Dataset} that wraps the user's \texttt{Dataset} (which will be hereby referred to as the \textit{sub-\texttt{Dataset}}), and caches samples as they are requested from the \texttt{Dataset}. For instance, if the $42^{nd}$ sample is requested from the cache \texttt{Dataset}, it will check MongoDB to see if the $42^{nd}$ sample has been cached this training session. If it has, it will be returned to the caller. If not, the \texttt{Dataset} will retrieve the $42^{nd}$ sample from the sub-\texttt{Dataset}, then cache it in MongoDB (if desired) and return it to the caller. If the cache is full before the sample is added, then existing samples are evicted following a first-in-first-out policy.

\subsection{Pre-fetching}

\begin{figure}[t]
    \centering
    \includegraphics[width=\columnwidth]{\altimages{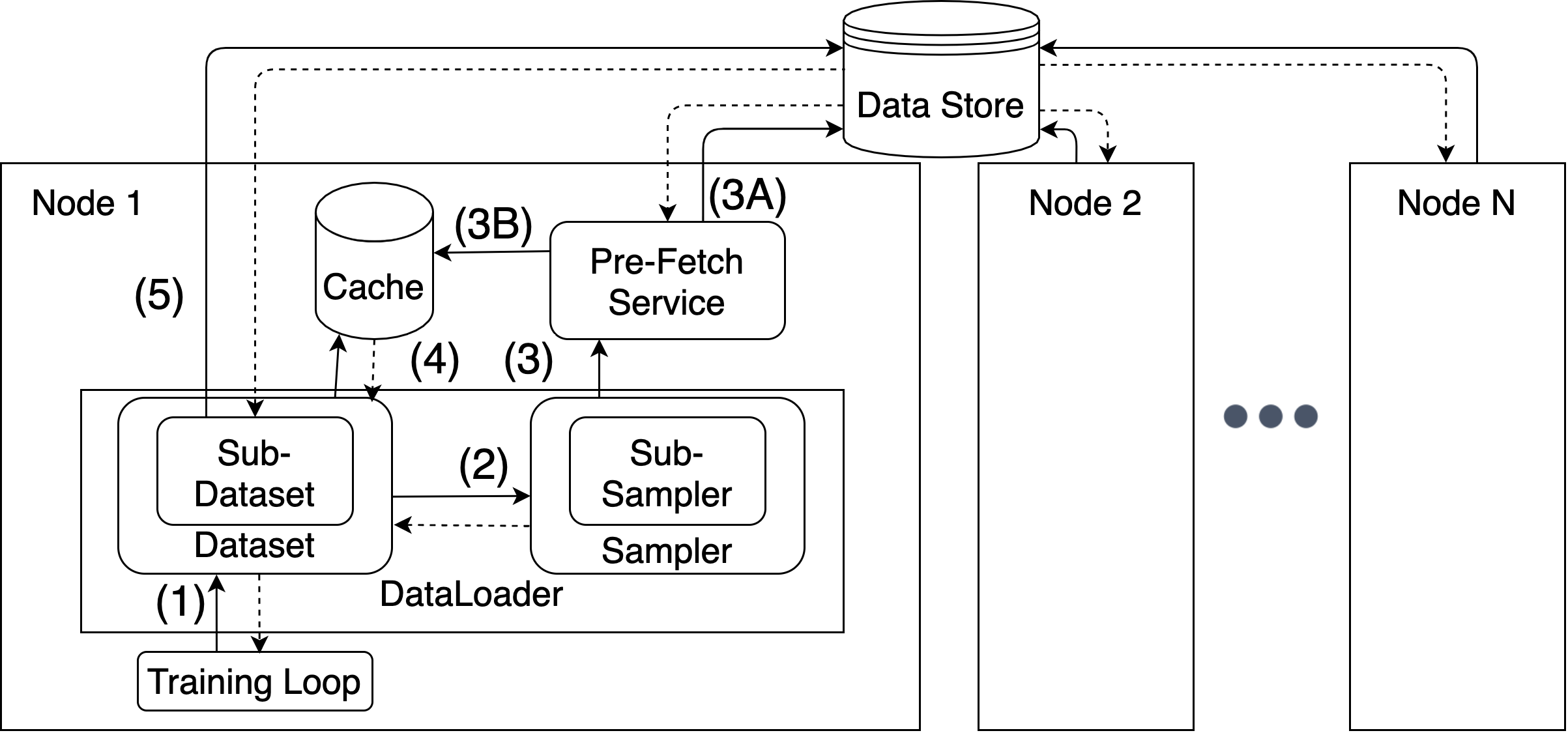}{assets/data_flows/data_flows_labeled_large.png}}
    \caption{A visualization of the pre-fetching and caching mechanism. The numbered lines show the order of data flow. Notably, steps 3A and 3B are asynchronous and may happen any time after step 3.}
    \label{fig:diagram_data_flows_labeled}
\end{figure}

To better overlap computation and communication, we decided to decouple the fetching process from the training loop as much as possible by using a separate service to pre-fetch the data. This service retrieves samples from the bucket in parallel, and once they are all ready, they are cached in parallel. Figure \ref{fig:diagram_data_flows_labeled} outlines this system.

We subclass the \texttt{Sampler} to integrate with PyTorch's API more easily. Our \texttt{Sampler} serves as a wrapper for another \texttt{Sampler}---the \textit{sub-\texttt{Sampler}}. The sub-\texttt{Sampler} can be any user-provided \texttt{Sampler}; the use of our wrapped \texttt{Sampler} does not require altering the order in which samples are read. During its operation, the \texttt{Sampler} requests one fetch size worth of samples from the sub-\texttt{Sampler} and returns these to the \texttt{DataLoader} transparently. Internally, this is simply a queue of the next indices that the sampler will provide.

When the \texttt{Sampler} requests new samples from the sub-\texttt{Sampler}, it also sends a request to the pre-fetch service. Upon receiving this request, the pre-fetch service immediately sends a response and spins up a subprocess to cache the samples from the storage bucket in the background.  This not only prevents the training loop from waiting for the samples to be cached, it also enables the service to always be ready for new requests from the \texttt{Sampler} without having to wait for a cache operation to complete. Additionally, to side-step the inability to perform batch-downloads mentioned in Section \ref{section:moving-to-cloud}, we simulate a batch download by downloading multiple files in parallel. After the training loop consumes enough samples from the queue, the pre-fetch threshold will be reached, at which point the pre-fetch service requests another round of samples. The number of samples fetched is still the fetch size, no matter the number of indices remaining in the queue.

\begin{figure}[t]
    \centering
    \includegraphics[width=\columnwidth]{\altimages{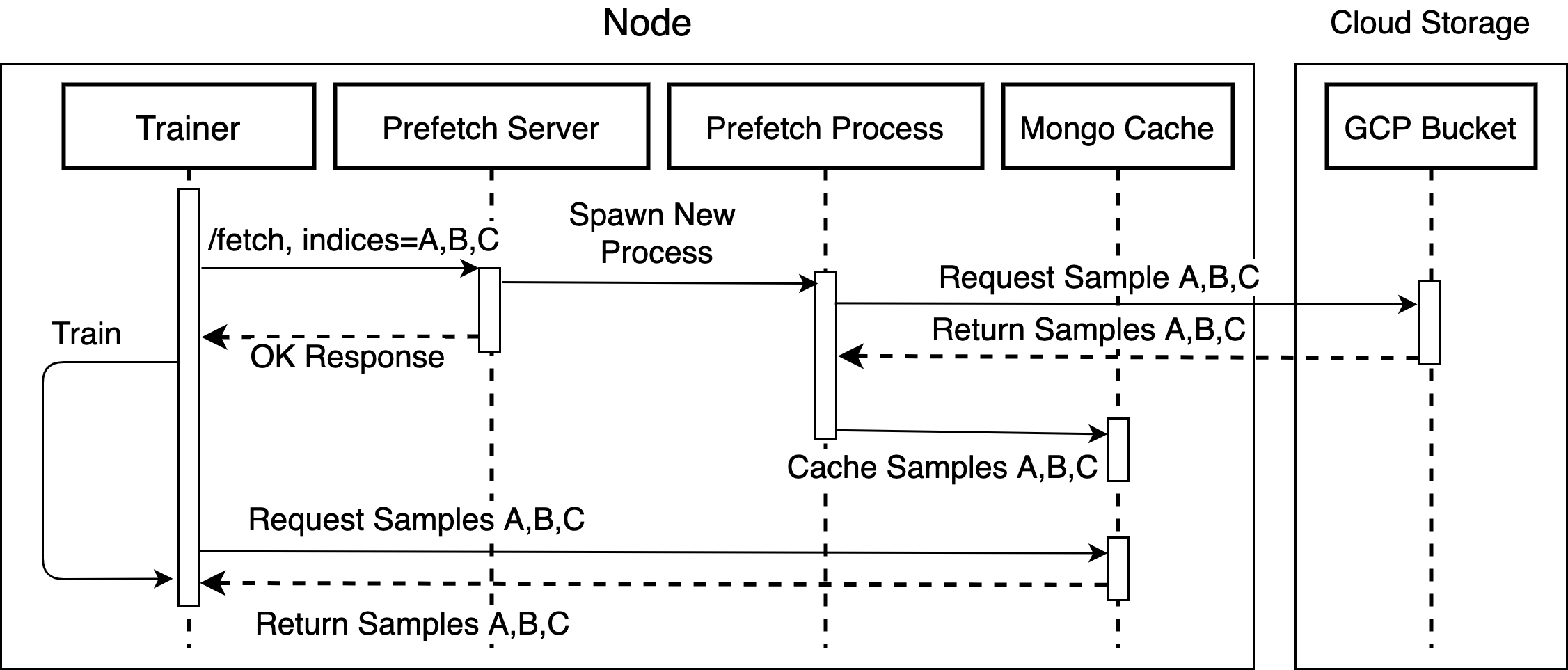}{assets/Prefetch_Sequence_diagram_large.png}}
    \caption{A sequence diagram showing the flow of data during pre-fetching. This shows one possible scenario, in which the training loop requests samples after all of them have been cached.}
    \label{fig:prefetching_sequence_diagram}
\end{figure}

Figure \ref{fig:prefetching_sequence_diagram} shows a sequence diagram for one possible scenario in which the samples requested are all cached before they are requested by the training loop.
There is no guarantee that this will be the case, though. Just as easily, the caching could have completed after the training loop asked for those samples, in which case the \texttt{Dataset} would retrieve the samples from the GCP bucket. To ensure time is not wasted, we choose to not have the worker perform a cache insert in this case, as the pre-fetch service will eventually perform this insert operation. Our choice to leave the \texttt{DataLoader} in the dark about the status of the pre-fetching can leave the system vulnerable to repeated cache misses if the pre-fetch service falls behind the training loop. In exchange, however, by not adding more points of communication, we allow more overlap between data loading and training.

\section{Results}
\label{chapter:results}

\subsection{Experiment Methodology}
    \label{section:experiment-setup}
    These results quantify the training time efficiency of our caching and pre-fetching training workflow for a cloud-based training environment.

\para{Setup.} Since this research concentrates on exploring data parallel distributed deep learning, the experiments use a three-node distributed data parallel setup, all in Google's us-east-1c data center. Evaluations were only done in this three-node configuration, as each node in \sysname operates independently during the data loading phase. Indeed, our preliminary testing demonstrated that a three-node setup spent less time loading data than a single-node setup, solidifying our decision to focus on this configuration.

Each node consists of a Google Compute Engine VM configured with 2 Intel Haswell vCPUs, 13 GB of memory, and one NVIDIA Tesla K80. The K80s were selected for three reasons. First, they are a popular choice among other researchers \cite{shijian-li2020characterizing, kung-fu}. Second, our evaluation is strictly concerned with the amount of time spent getting data loaded for the GPU to use, not the actual time spent training; this makes the choice of GPU not of material concern. Last, K80 GPUs are one of the cheapest options available on Google Cloud Platform \cite{gcp_gpu_pricing}, which made them a fitting choice given the lower importance of our GPU choice.

These nodes train with two workloads: CIFAR-10 on ResNet-50, and MNIST on a CNN with two convolutional layers and a single fully-connected layer. To implement data parallelism, each model is trained on an evenly-sized partition of the data generated by PyTorch's \texttt{DistributedSampler} to produce a random subset every epoch. This was selected due to its recommendation by the PyTorch documentation as a sampler for distributed data parallel workloads \cite{website:PyTorch_data}, but by no means is our methodology limited to such a partitioning scheme.

\para{Metrics.} We choose two metrics, \emph{data loading time} and \emph{cache miss rate} and measure both of these for the first and second epochs. Specifically, the miss rate is calculated as a proportion of the number of samples the worker did not find in the cache out of the number of samples requested in the epoch. In addition, the data loading time is the time taken to load an epoch's samples from a given storage medium, whether it's the cache, disk, object storage, or a combination thereof. These results include measurements for only the first two epochs because preliminary experimentation revealed that using this paper's approach, the second epoch approximates the performance of every subsequent epoch in terms of data loading time and cache miss rate. This mirrors other research that indicates that per-epoch training time remains relatively consistent after an initial warm-up period \cite{shijian-li2020characterizing, pinto2018hoard}. We report average data loading time and cache miss rate over three nodes and then over three trials.

Referring back to the data flow shown in Figure \ref{fig:diagram_data_flows_labeled}, data loading time includes all time spent between the \texttt{Dataset} and the cache, and the sub-\texttt{Dataset} and the data store (steps 4 and 5 on the diagram respectively). This is the time taken to attempt to load a sample from the cache and to load it from the bucket as a fallback.

\para{Data Loading Baselines.} We compare our prototype that uses caching and pre-fetching for cloud bucket-based training against a set of baselines, each of which uses our three-node distributed data parallel setup. The first baseline trains the model by pulling training data from disk, without relying on external services. The second baseline fetches all samples from GCP buckets without any caching or pre-fetching (i.e.\ skipping steps 2-4 in Figure \ref{fig:diagram_data_flows_labeled}). This helps measure the total amount of training time that our system saves compared to naïvely assuming the network will be fast enough to retrieve the training data. The third baseline fetches data from GCP buckets, but introduces the cache layer without the pre-fetching layer. We repeat this at various cache sizes to quantify how much of the system's benefit comes from caching and how much comes from pre-fetching.

    \subsection{Summary of Key Findings}

    \begin{figure}[t]
        \centering
        \includegraphics[width=\columnwidth]{\altimages{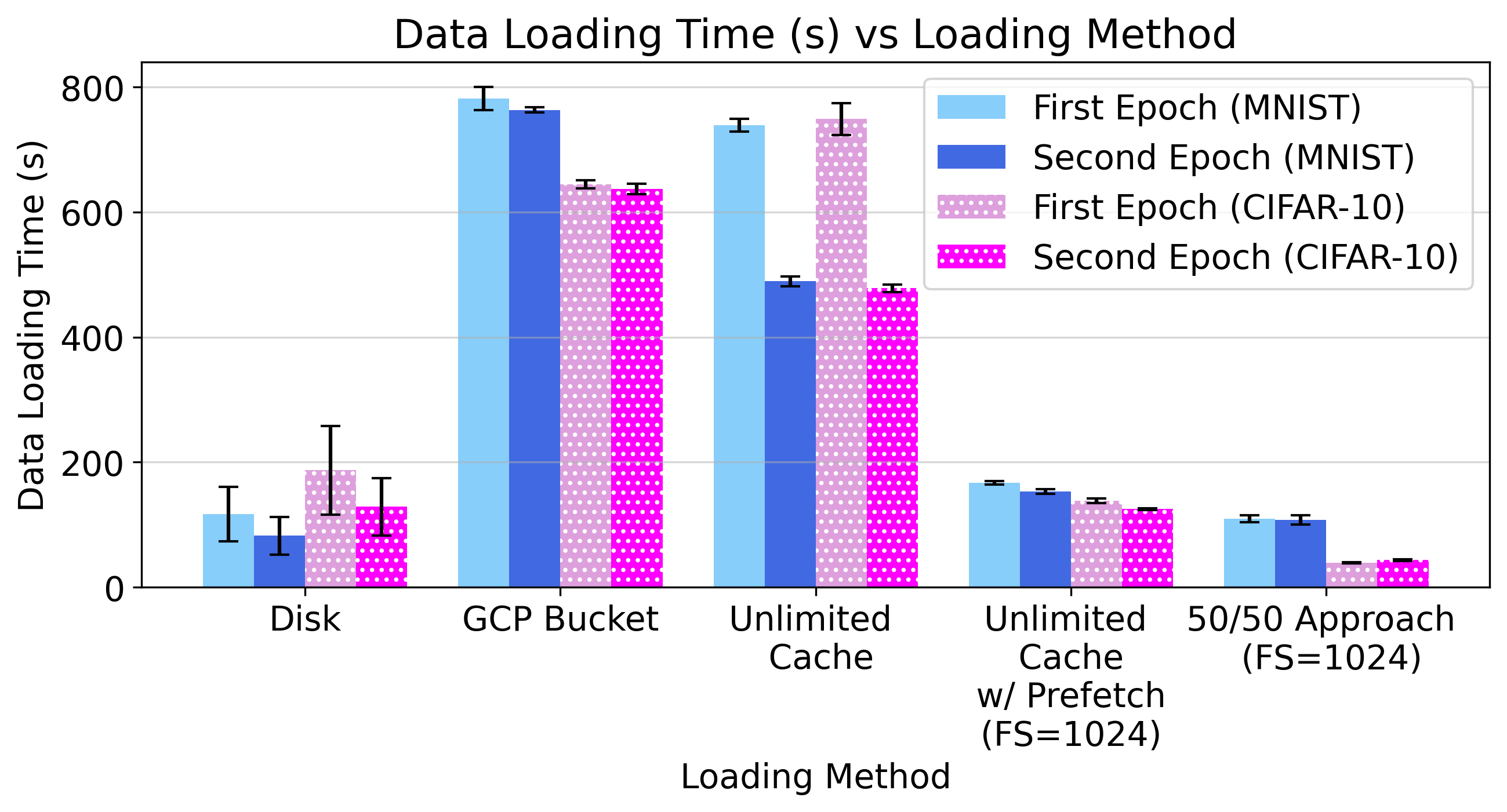}{assets/low_res/combined/baselines.png}}
        \caption{Data loading time comparison. We show that \sysname, when configured properly (i.e.\ 50/50 approach), can achieve near-disk data loading time.
        }
        \label{fig:graph_baselines}
    \end{figure}
    \begin{figure}[t]
        \centering
        \includegraphics[width=\columnwidth]{\altimages{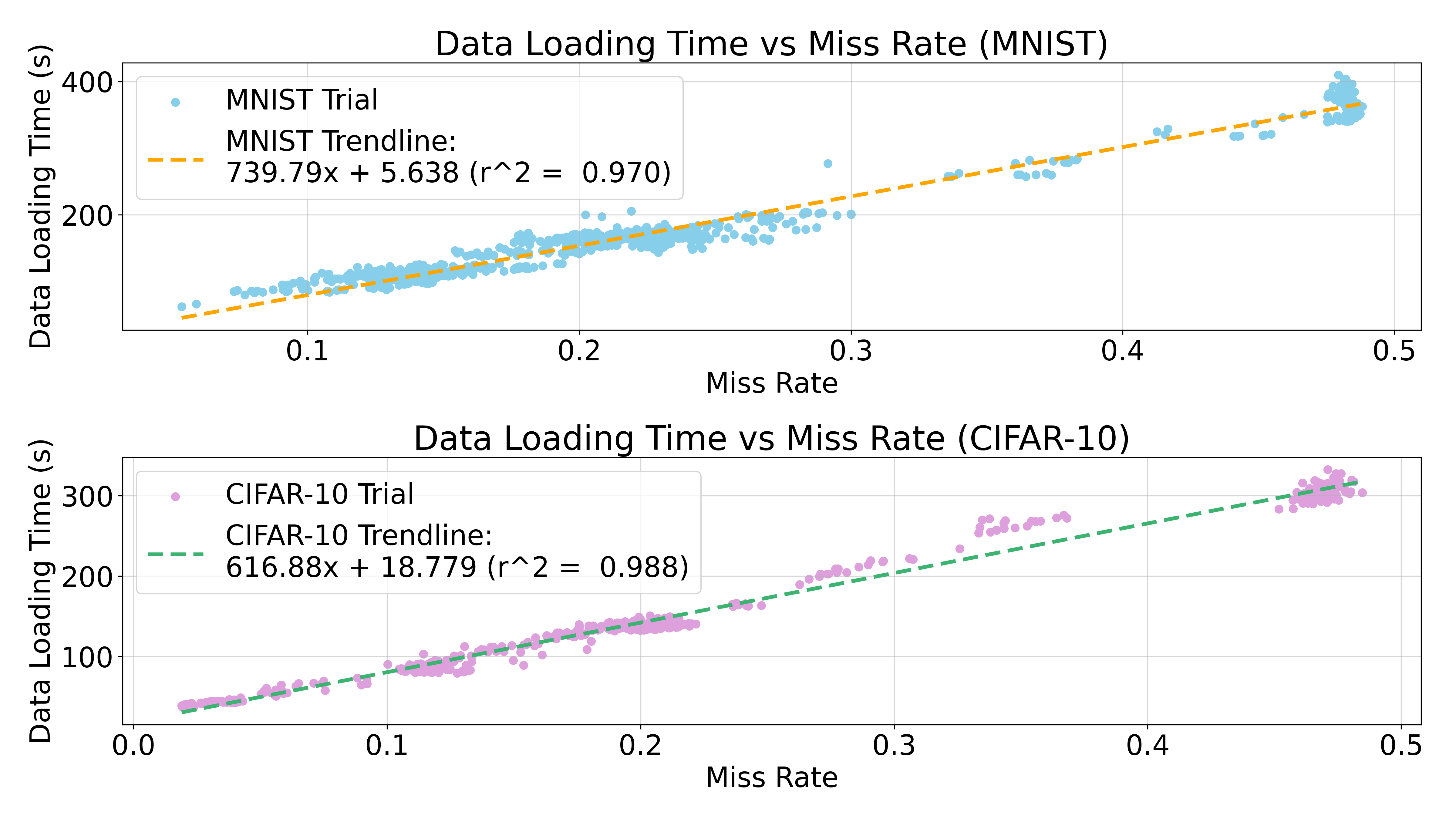}{assets/low_res/combined/Data Loading Time vs Miss Rate Updated Squished and Titles.png}}
        \caption{The linear relationship between miss rate and data loading time for both MNIST and CIFAR-10 trials. \note{Figure has been updated to show more datapoints, separating out MNIST and CIFAR-10 trials}}
        \label{fig:graph_dl_miss_rate}
    \end{figure}

    Our evaluation demonstrates that \sysname achieves significant speedup of data loading time compared to naïvely loading data from object storage, for both training workloads. Specifically, we observed that, compared to reading from GCP buckets directly, the \emph{50/50 approach} (described shortly) reduced data loading time by 85.6\% and 93.5\% for MNIST and CIFAR-10, respectively. Figure~\ref{fig:graph_baselines} summarizes the data loading time using different approaches. Measuring the amount of time that the training loop spends waiting for data reveals that loading data from object storage results in an increase in loading times between 8 and 16 times when compared to loading data from disk. In our testing, this penalty ended up being a significant portion of the time spent, as our models spent an average of 14.7~s and 147.2~s training on MNIST and CIFAR-10 respectively: a figure that is dwarfed by the several-hundred second data loading time. One might wonder why the GCP loading time is lower for CIFAR-10. This is simply because in our testing, there were fewer samples in the partition for CIFAR-10 than for MNIST. By plotting the results of our individual trials in Figure \ref{fig:graph_dl_miss_rate}, we can see that data loading time changes linearly with miss rate.\footnote{It is worth noting that this plot only includes trials that used both pre-fetching and caching, as the trend is more easily seen on a smaller plot. However, this trend does hold for higher miss rates than 0.5.} As such, this allows us to account for this difference in dataset size by comparing miss rates, rather than absolute times.

    Adding a cache can net some improvement once the data is cached, but there still is a time penalty associated with getting the data into the cache to begin with. Dramatic gains arise once the pre-fetching solution is added to the mix, cutting loading times to be just double of the disk loading time; this can be lowered even further with some adjustments to fetch thresholds. These observations will be discussed in more detail in their respective sections.

    The most interesting and practical takeaways from our results are how they allow users to achieve performance much closer to that of loading training data from disk than that of loading data from a central data store alone, while still not persistently storing the training data on disk. We have found that it is best to take what we call the \emph{50/50 approach}: setting both the fetch size and the pre-fetch threshold to half the cache size. This effectively ensures that one fetch of data is always cached and that each new fetch fits fully within the cache. Figure \ref{fig:graph_baselines} shows that this configuration is the fastest of everything tested. We suspect that it beats our disk baseline in part because MongoDB caches frequently accessed entries in memory. The best settings we found work well across different workloads tested, suggesting that these properties are fairly general characteristics of our system. The potential cost-saving effects of this are discussed in Section \ref{subsection:cost}.

\subsection{Effect of Caching On Data Loading Time}

    We first evaluate the effectiveness of caching alone to help determine how necessary each component of our system is in improving data loading time. These experiments use the aforementioned three-node setup, partitioning scheme, and workloads. As shown in Figure \ref{fig:graph_baselines}, an unlimited cache can have some benefits over fetching directly from the object store during the second epoch. To determine how effective a cache is at more realistic sizes, an unlimited cache is tested against caches that can only hold 25\%, 50\%, and 75\% of the samples a node would be assigned in one partition.

    \begin{figure}[t]
        \centering
        \includegraphics[width=\columnwidth]{\altimages{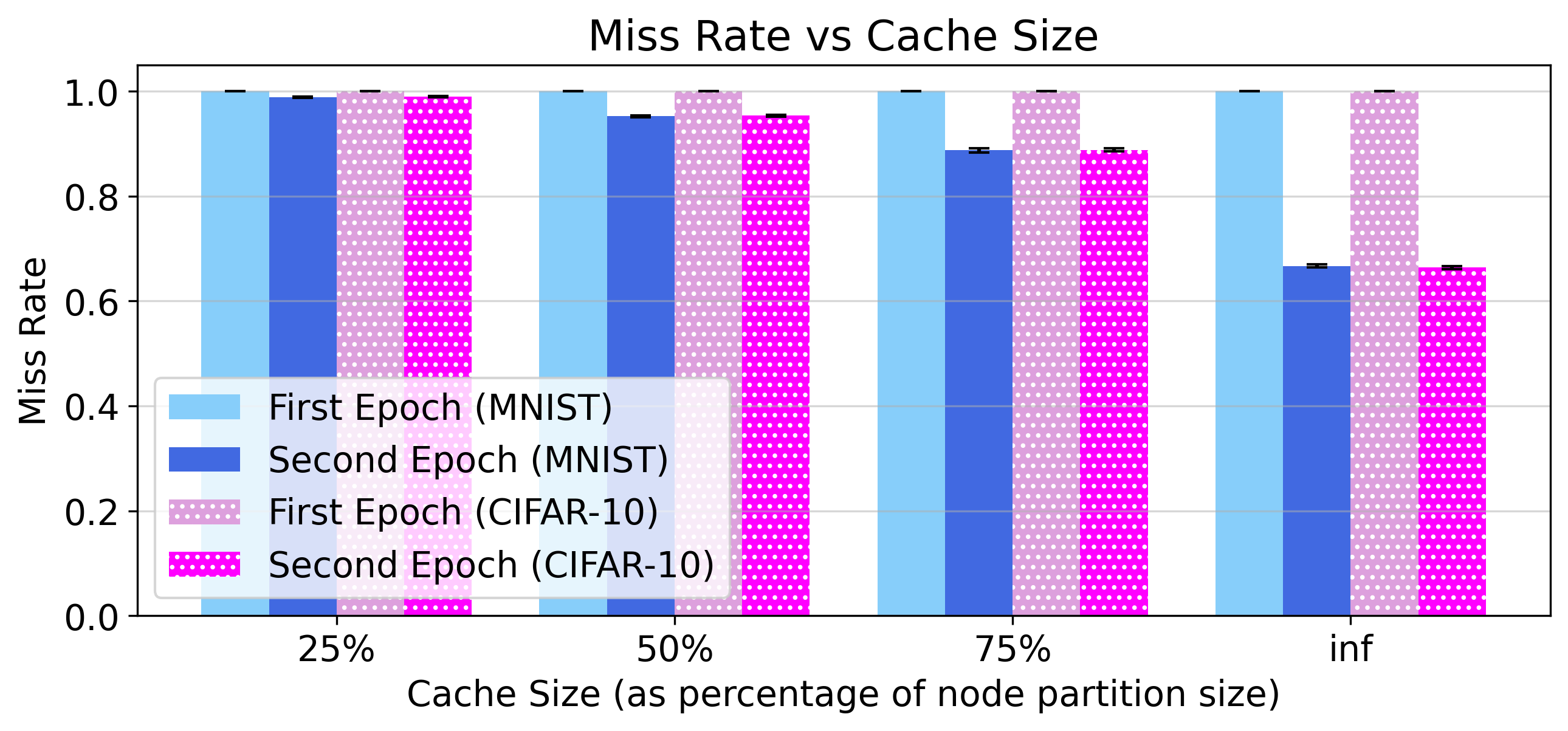}{assets/low_res/combined/Combined Baseline Miss Rate vs Cache Size.png}}
        \caption{The effect of cache size on miss rate for both CIFAR-10 and MNIST for an unlimited cache against three proportions of the partition size.}
        \label{fig:graph_unlimited_cache_miss_rate}
    \end{figure}

    Our results show that caching alone is not particularly well-suited to the data parallel case our research explores.  Under these circumstances, even a cache large enough to hold the entire dataset does not show impressive gains within the first two epochs. As Figure \ref{fig:graph_unlimited_cache_miss_rate} demonstrates, the miss rate for an unlimited cache's second epoch is very high at 66\%, which makes sense given that the partition across three nodes is random every time.\footnote{Although training over many epochs would eventually cache the entire dataset, our testing focuses on improvements within the first two epochs.} Each node should be able to cache the entirety of whatever random third of the data it was assigned in the first epoch partition. When the data is re-sampled for the second epoch, about a third of these samples are likely to be reassigned to that node, which would result in a 33\% hit rate and a 66\% miss rate, aligning very closely with what we observed. The fact that these numbers are so consistent across workloads further underscores this point: the miss rates are probabilistic rather than dataset-dependent.

    Figure \ref{fig:graph_unlimited_cache_miss_rate} furthermore demonstrates that the miss rate climbs rapidly as the cache size is constrained. This is especially troublesome because a constrained cache better reflects the situations in which our system is designed to help---where the dataset cannot feasibly be held on each node. If there is enough disk space to cache the entire dataset, that space would be better used to hold the dataset directly, rather than fetching it from object storage. Limiting the cache size to even 75\% of the data partitioned to a node results in about a 90\% miss rate. Clearly, caching alone does not sufficiently counteract the penalty of accessing data from the object store.

    \textbf{Summary:} Caching can help improve data loading times, but caching alone is not a robust strategy, especially when applied with partitioned data. The benefits of caching diminish if the cache is not large enough to hold the entire dataset.

\subsection{Maximizing Cache Benefits with Pre-Fetching}
\label{section:prefetch-results}

    \begin{figure}[t]
        \centering
        \includegraphics[width=\columnwidth]{\altimages{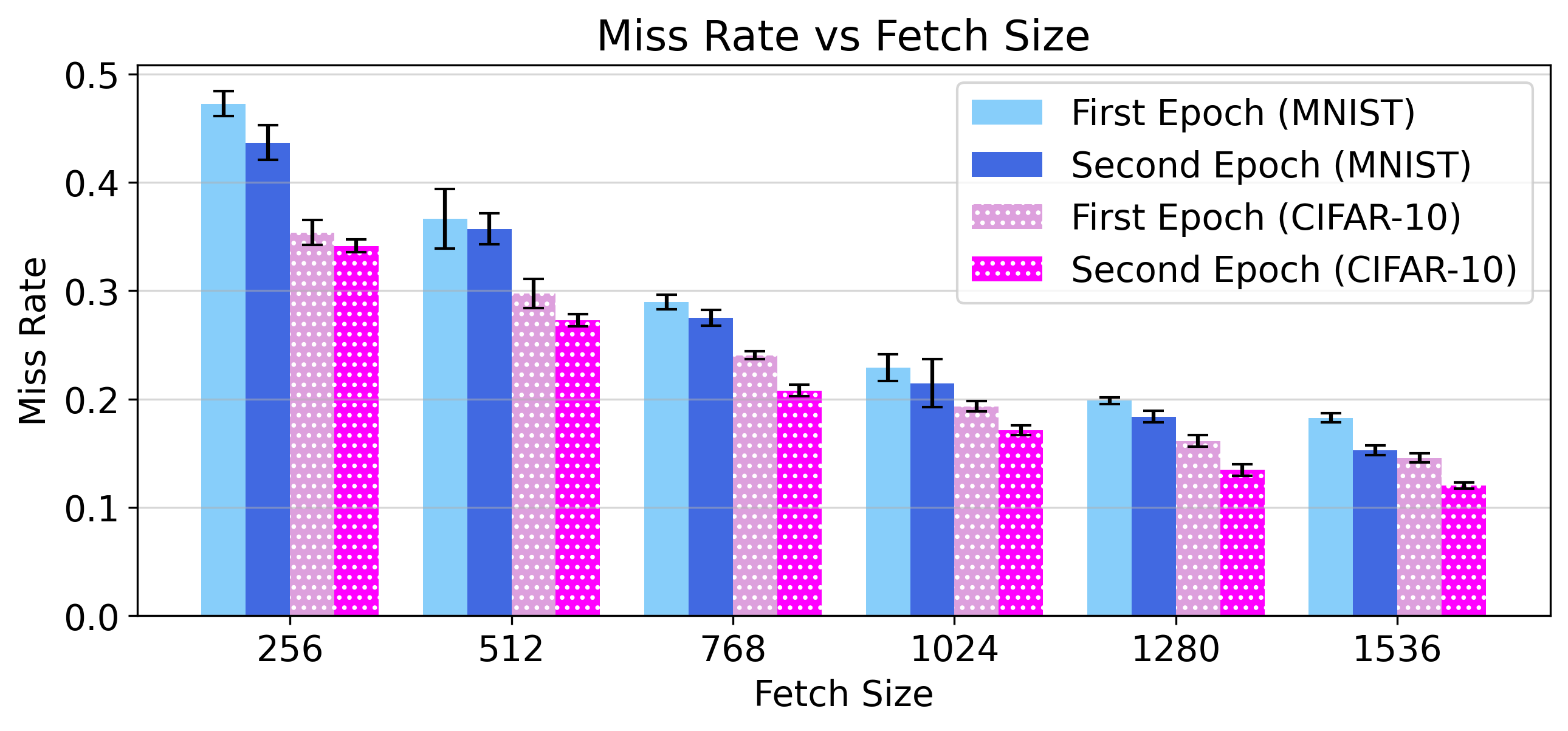}{assets/low_res/combined/Combined Miss Rate vs Fetch Size.png}}
        \caption{The effect of increasing the fetch size on miss rate. In general, increasing the fetch size will decrease the miss rate.}
        \label{fig:graph_miss_rate_fetch_size}
    \end{figure}

    Building on top of the caching approach, pre-fetching further reduces data-loading time by strategically pre-populating the cache. To evaluate this approach, we left the cache unlimited and varied the fetch size in increments of 256 samples. Figure \ref{fig:graph_miss_rate_fetch_size} shows the clear benefits of increasing the amount of data that is fetched ahead of time.

\begin{figure}[t]
        \centering
        \begin{subfigure}{\columnwidth}
			\includegraphics[width=\textwidth]{\altimages{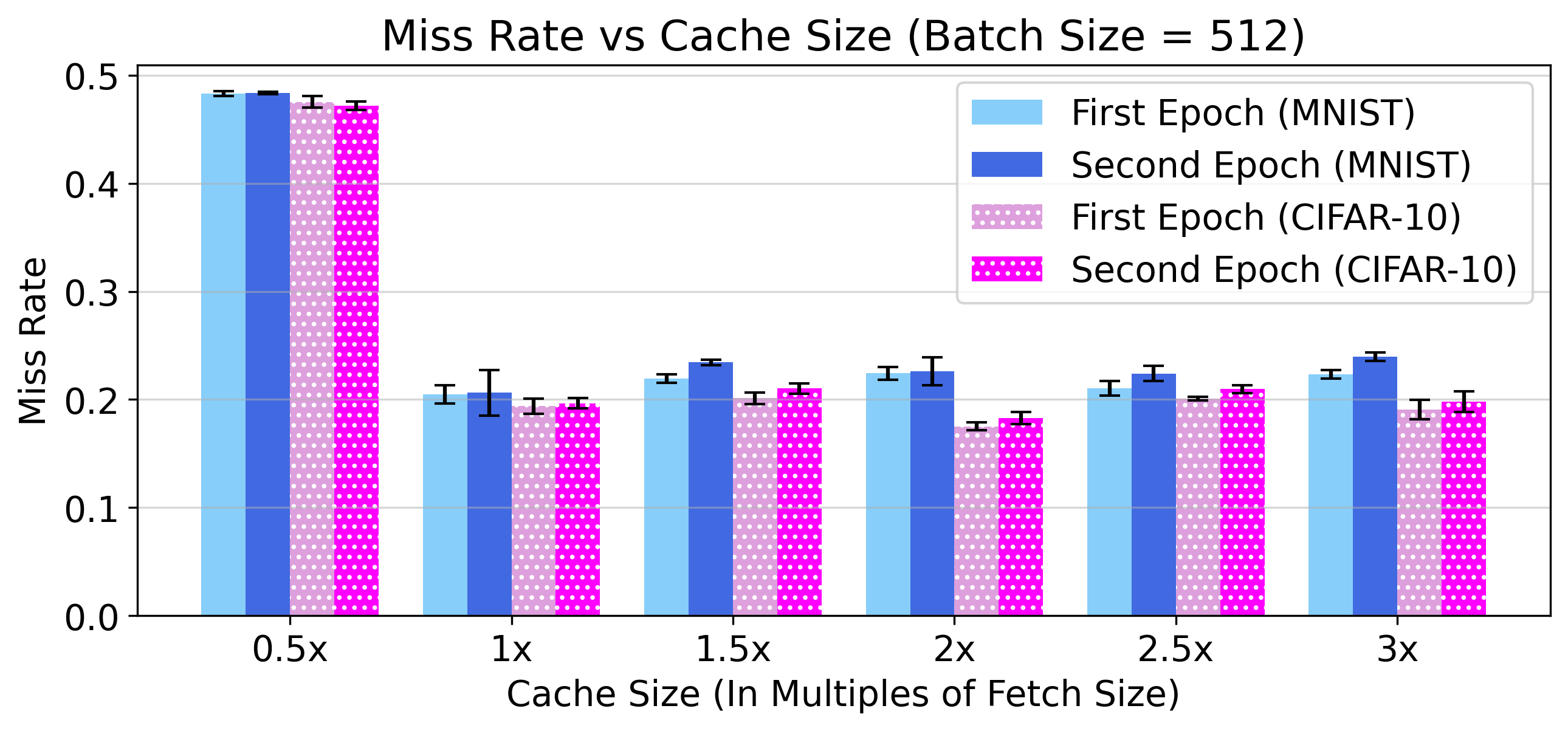}{assets/low_res/combined/Combined Miss Rate vs Cache Size (bs=512).png}}
        \end{subfigure}
        \begin{subfigure}{\columnwidth}
            \includegraphics[width=\textwidth]{\altimages{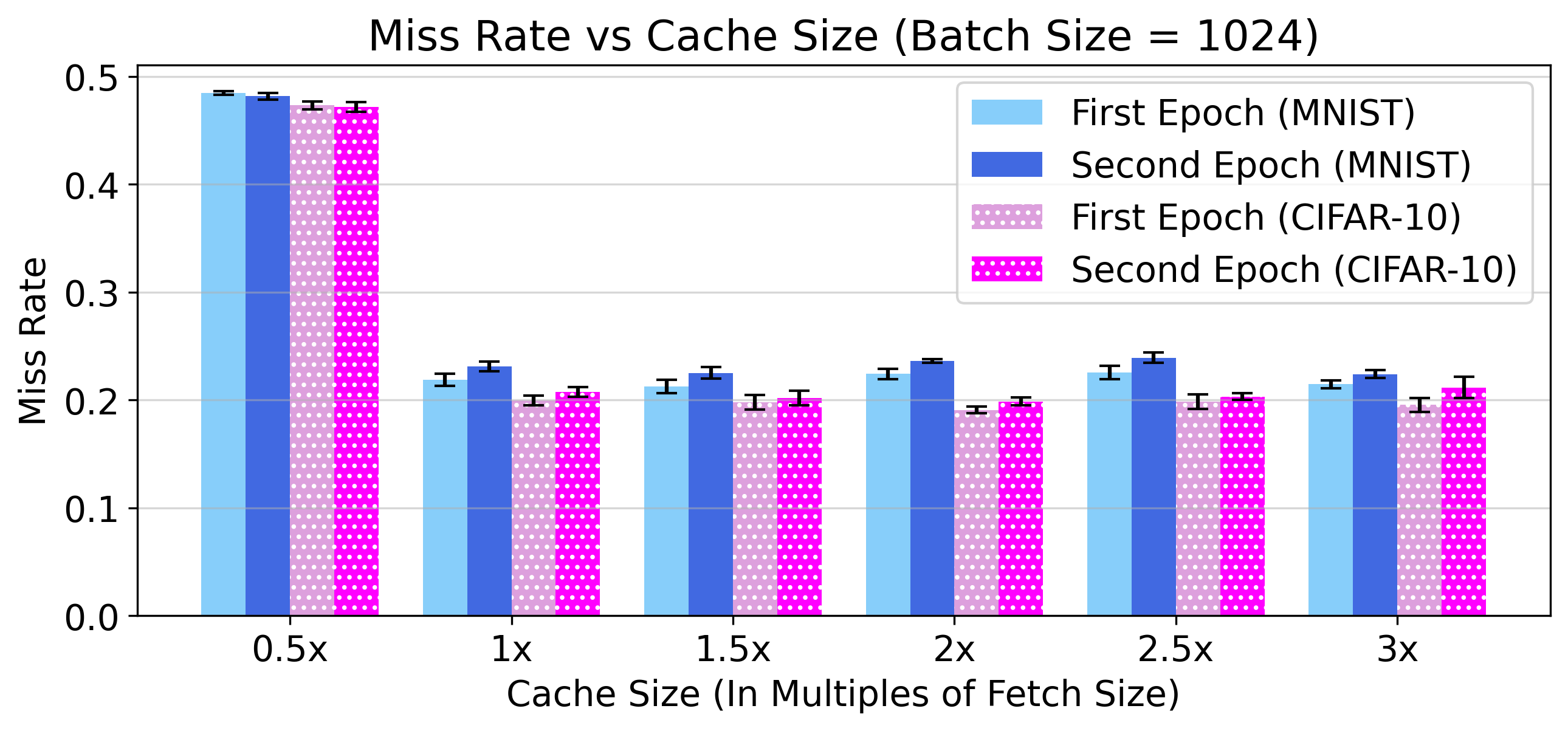}{assets/low_res/combined/Combined Miss Rate vs Cache Size (bs=1024).png}}
        \end{subfigure}
        \caption{The effect of increasing the cache size with a constant fetch size. After 1x, the benefits of doing so vanish.}
        \label{fig:graph_cache_miss_rate}
    \end{figure}

   As mentioned previously, an unlimited cache is simply not practical. Our testing demonstrates, though, that if the cache size is limited, similar performance to the unlimited cache can be achieved for any cache size greater than or equal to the fetch size. To show this, we conduct an experiment in which the fetch size is fixed at 1024 samples (i.e.\ less than 10\% of each node's partition for both CIFAR-10 and MNIST) and the cache size is varied. Figure \ref{fig:graph_cache_miss_rate} shows the result of this experiment; after the cache size exceeds the fetch size, the difference in miss rate becomes negligible. The plot for a batch size of 1024 underscores the significance of this result: with a fetch size equal to the batch size, maximum performance can be obtained by \textit{storing only the batch we are currently training on}.

    Curiously, the miss rate of the CIFAR-10 trials is consistently lower than the MNIST trials. This can be attributed to the large difference in compute time between our custom convolutional network and ResNet. On average, the compute time of ResNet was 15x slower than the convolutional network. This makes sense, given that ResNet is a far more complex model. During this training time, the pre-fetcher has ample opportunity to download more data, resulting in a lower miss rate for CIFAR-10.

    Recall from Section \ref{chapter:implementation} that the default pre-fetching approach only fetches new samples when the \texttt{Sampler}'s queue of samples has been depleted (i.e.\ a pre-fetch threshold of zero). We run experiments to see how adjusting the pre-fetch threshold affects data loading time. Again holding the fetch size constant at 1024, we adjust the cache size to different multiples of the fetch size (0.5x, 1x, 2x, 3x). For each of these cache sizes, we test a pre-fetch threshold of 25\%, 50\%, and 75\% of the cache size.

    \begin{figure}[t]
        \begin{subfigure}{\columnwidth}
            \centering
            \includegraphics[width=\columnwidth]{\altimages{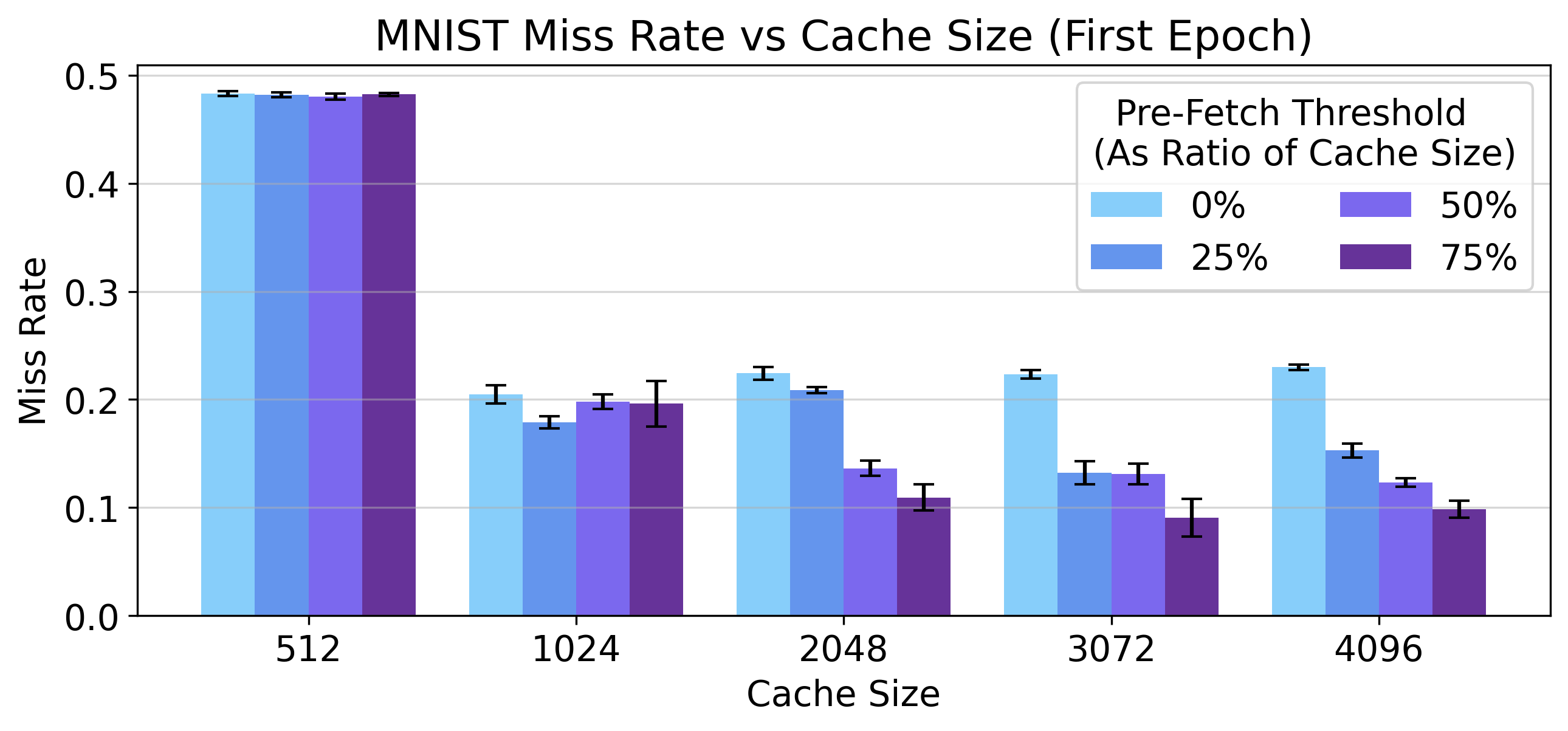}{assets/low_res/mnist/mnist mqs miss rate vs cache size (first epoch).png}}
			\includegraphics[width=\columnwidth]{\altimages{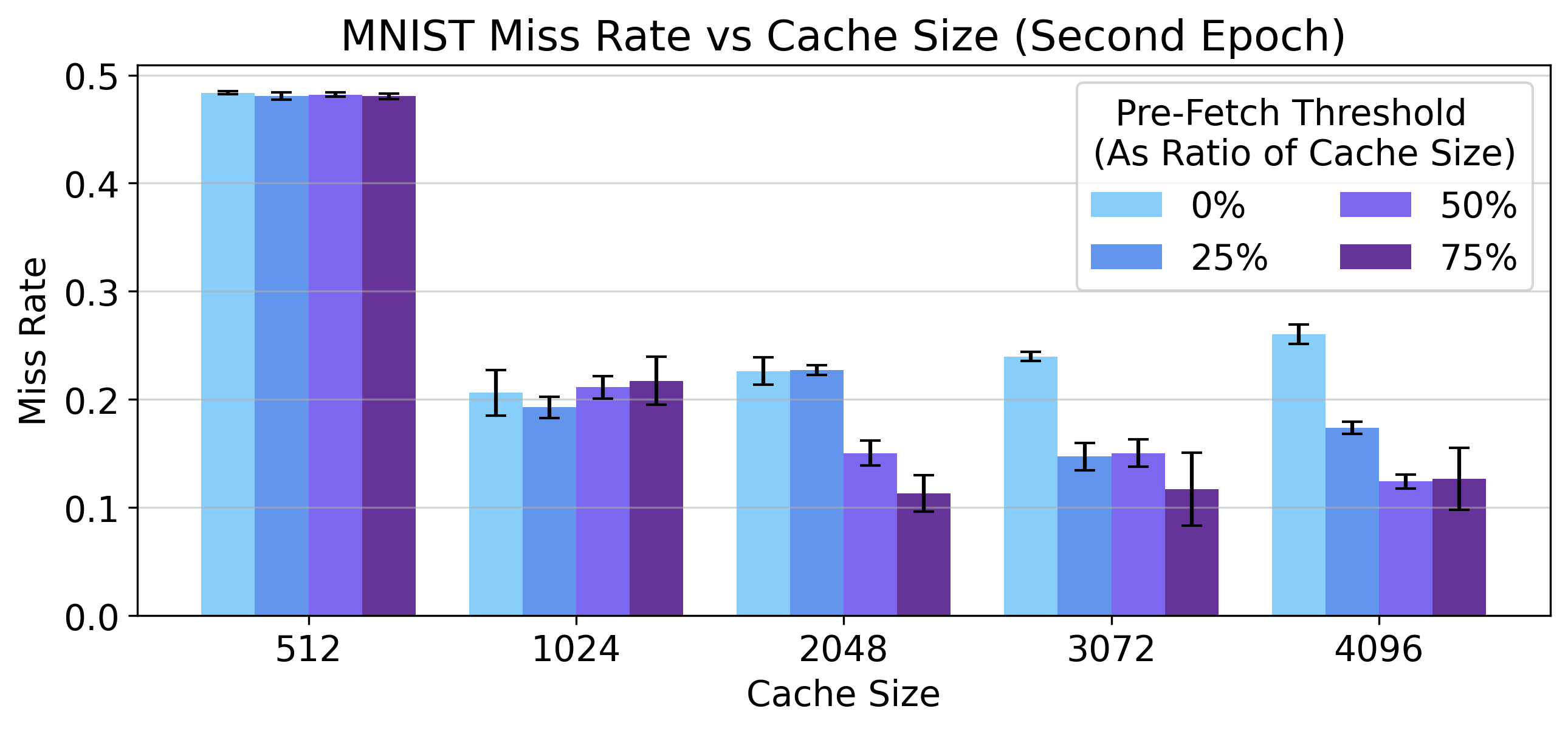}{assets/low_res/mnist/mnist mqs miss rate vs cache size (second epoch).png}}
            \caption{Comparing the effect of different pre-fetch thresholds on the MNIST workload.}
        \end{subfigure}
        \begin{subfigure}{\columnwidth}
            \centering
            \includegraphics[width=\columnwidth]{\altimages{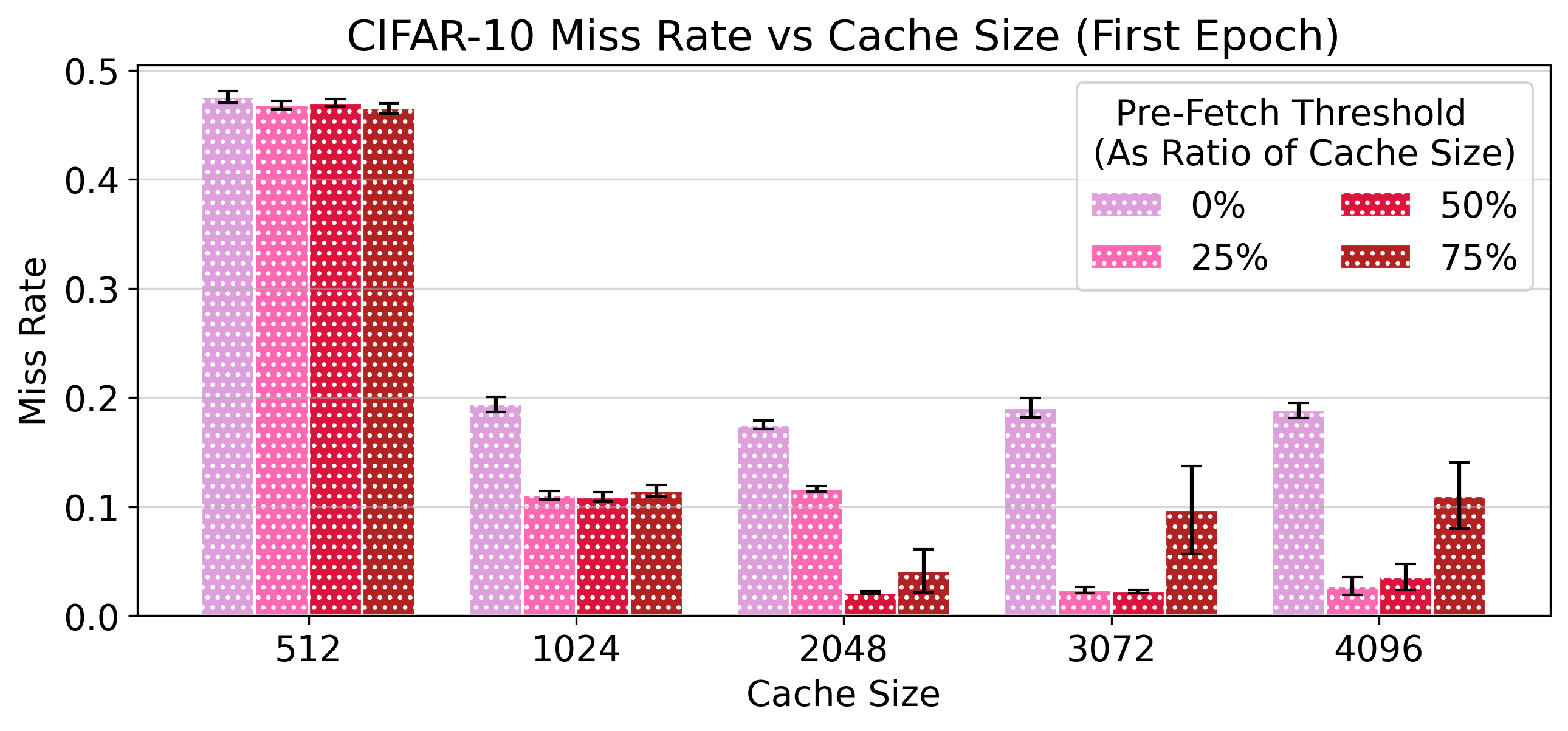}{assets/low_res/resnet/RESNET MQS Miss Rate vs Cache Size (First Epoch).png}}
            \includegraphics[width=\columnwidth]{\altimages{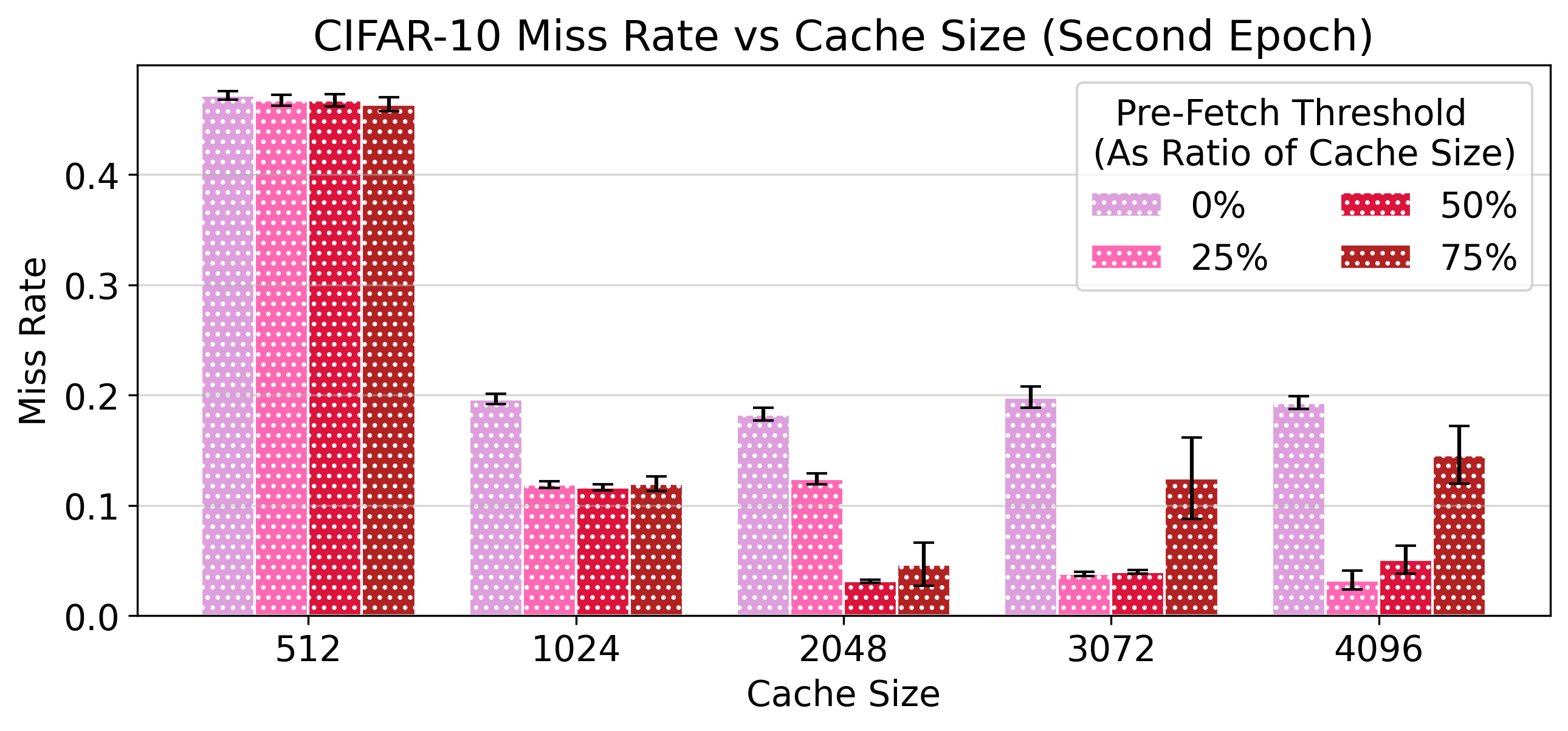}{assets/camera-ready/RESNET MQS Miss Rate vs Cache Size (Second Epoch).png}}
            \caption{Comparing the effect of different pre-fetch thresholds on the CIFAR-10 workload.}
        \end{subfigure}
        \caption{Comparing the effect of different pre-fetch thresholds (as a proportion of cache size) on cache miss rate for various cache sizes across epochs and models.}
        \label{fig:min_queue_size}
    \end{figure}

    Figure \ref{fig:min_queue_size} shows that with a cache size of 2048 and a pre-fetch threshold of 1024 (i.e.\ 50\%), we can see a 31\% and 80\% decrease in miss rate for MNIST and CIFAR-10 respectively, with diminishing returns afterwards. In other words, maximum benefit can be obtained by setting the fetch size to half of the cache size, and setting the pre-fetch threshold equal to the fetch size. This means that the pre-fetcher will prepare a full fetch of data just as there is exactly one fetch worth of data remaining, referred to earlier as the 50/50 approach. This produces a similar effect to doubling the fetch size. Although the magnitudes of the miss rate differ between the two workloads, the trend holds across them. This difference in magnitude can again be attributed to the difference in compute time.

    Although setting the pre-fetch threshold to 75\% of the cache size may appear advantageous in some cases (e.g. a cache size of 2048 on MNIST), this does not seem to generalize well. As the CIFAR-10 graphs reveal, a pre-fetch threshold of 75\% does not always show the same gains as a threshold of 50\%. Furthermore, the error bars on the 75\% pre-fetch thresholds on both workloads indicate a lack of reliability; the other thresholds proved to be more reliable. Similarly, though there are some cases where the 25\% pre-fetch threshold edges out the performance of a 50\% pre-fetch threshold, this is hardly consistent, and not worth considering compared to the more reliable 50\% threshold.

\begin{figure}[t]
        \centering
        \includegraphics[width=\columnwidth]{\altimages{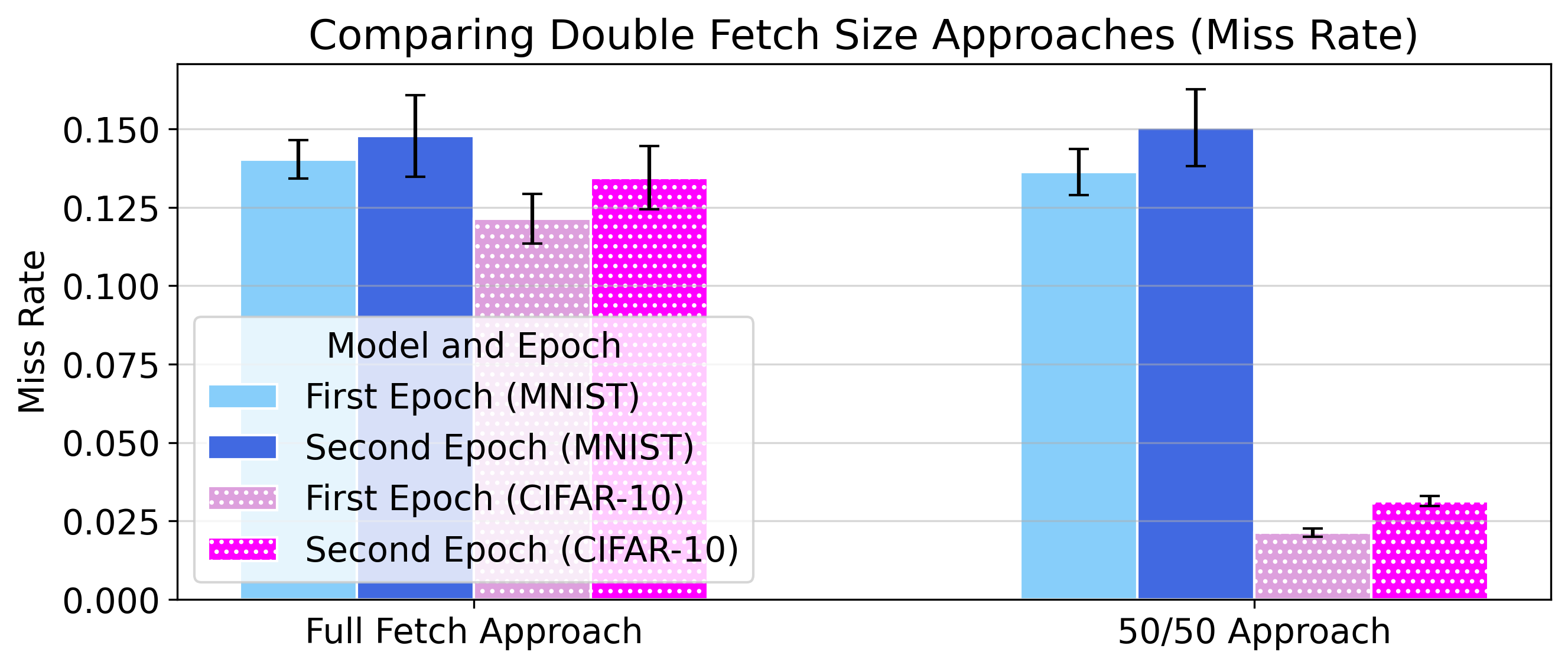}{assets/low_res/combined/Comparing Double Fetch Size Approaches (Miss Rate).png}}
        \caption{Comparing the best-found settings for fetch size and pre-fetch threshold across models for a cache size of 2048.
        }
        \label{fig:graph_comparing_double_fetch_size_approaches}
    \end{figure}

    Given the choice between doubling the fetch size (termed \emph{Full Fetch}) and making the fetch size and pre-fetch threshold equal to 50\% of the cache size, one should prefer the latter. Figure \ref{fig:graph_comparing_double_fetch_size_approaches} demonstrates this; though for MNIST the difference is quite small, applying the 50/50 approach yields an 83\% decrease in miss rate for CIFAR-10. Therefore, we recommend using the 50/50 approach for more computationally-intensive workloads (i.e.\ with longer per-epoch computation time), as the difference can be quite drastic, even beating our disk baseline, as shown in Figure \ref{fig:graph_baselines}. The speedup for the CIFAR-10 workload can be in part attributed to MongoDB's in-memory caching mechanism. To validate that this is the case, we examine MongoDB's internal cache metrics\footnote{Specifically, \texttt{db.serverStatus().wiredTiger.cache["bytes currently in cache"]} was checked.}; they indicate that the node's entire partition is in memory after the first epoch, though this may not be true with larger datasets.

    \textbf{Summary}: It is always beneficial to pre-fetch more data, but doing so increases the amount of disk space used. If the cache is limited, it does not need to be larger than the fetch size, but we can make the data loading time even lower by raising the threshold for the next fetch.

\subsection{Cost}
\label{subsection:cost}

\begin{table*}[t]
    \centering
    \caption{The modeled cost of training each workload for two epochs. Totals that are underlined save money compared to the disk baseline. \note{This table has been modified to show a cost breakdown.}}
\begin{threeparttable}
    \begin{tabular}{c | c | c | c | c || c | c | c | c }
         & \multicolumn{4}{c||}{\textbf{MNIST Costs}} & \multicolumn{4}{c}{\textbf{CIFAR-10 Costs}} \\
         \textbf{\textit{Method}} & \textit{API} & \textit{Storage} & \textit{Compute + Loading} & \textit{Total} & \textit{API} & \textit{Storage} & \textit{Compute + Loading} & \textit{Total} \\
         \hline
         \textbf{Baseline (Disk)} & n/a & \$1.95 & \$0.10 & \$2.05 & n/a & \$1.95 & \$0.28 & \$2.23 \\
         \textbf{Baseline (GCP)} & \$0.03 & \$1.93 & \$0.72 & \$2.68 & \$0.03 & \$1.93 & \$0.72 &  \$2.68 \\
         \thickhline
         \textbf{Full Fetch Approach (Fetch Size=1024)} & \$0.08 & \$1.93 & \$0.16 & \$2.17 & \$0.07 & \$1.93 & \$0.25 & \$2.25 \\
         \textbf{Full Fetch Approach (Fetch Size=2048)} & \$0.06 &  \$1.93 & \$0.12 & \$2.10 & \$0.05 & \$1.93 & \$0.23 & \underline{\$2.21} \\
         \textbf{50/50 Approach (Fetch Size=1024)}\tnote{\textdagger}{} & \$0.08 & \$1.93 & \$0.11 & \$2.12 & \$0.07 & \$1.93 & \$0.17 & \underline{\$2.17} \\
    \end{tabular}
    \begin{tablenotes}
    \item[\textdagger] \scriptsize This configuration uses as much cache space as Full Fetch Approach (FS=2048), which is twice the cache space of Full Fetch Approach (Fetch Size=1024).
    \end{tablenotes}
    \end{threeparttable}
    \label{table:cost}
\end{table*}

Despite our initial hypotheses, our approach did not universally lower the cost of running these workloads. Table \ref{table:cost} compares the monetary costs between \sysname with different configurations and two data loading baselines. Specifically, we highlight the cost of training from disk, the significant cost increase of moving to cloud buckets, and the cost of using our approach in comparison to those baselines. We produced these figures using the model described in Section \ref{section:cost-design}; the average data loading and compute times from these trials were used as \(t_c\) and \(t_d\), in lieu of estimated values.

To produce these figures, we model the cost of the baseline approaches and the best configurations of our approach. Using this model, we produc the results shown in Table \ref{table:cost}. The only scenarios where there's a cost saving is when using the full fetch approach with a fetch size of 2048, and the 50/50 approach with a fetch size of 1024. These gains are only realized on our ResNet/CIFAR-10 workload, which makes sense given that the model complexity allows the pre-fetching system to significantly reduce data loading time. In turn, this lowers the total epoch time  and therefore the ``compute + loading'' cost. It is worth noting that the costs from the per-request pricing scheme can quickly add up, which is a contributing factor to the figures in the table.

\textbf{Summary:} Though the bucket pricing model is cheaper per gigabyte than local storage, it is not always effective at lowering costs. Our caching and pre-fetching approach for bucket storage is favorable to storing samples on each node if the model compute time is sufficiently long.

\section{Discussion}
\label{chapter:discussion}

Our results show a clear performance improvement when using a caching and pre-fetching approach in a cloud-based distributed training environment. Specifically, by using \sysname, we can achieve near-disk data loading time when storing training data centrally in cloud bucket-based storage. Even though our approach does not lead to significant monetary savings as we had envisioned at the outset, we believe that \sysname is valuable for emerging training scenarios where it is not feasible to store training data on local disks.
For instance, serverless environments, by design, have very small amounts of temporary storage \cite{Cirrus, website:lambda_storage}. These constraints lend themselves well to using a bucket for storage. In addition, some learning environments store real-time data from the edge to train machine learning models, and will store this data in storage buckets \cite{alipour:microservice-logs, ma:medhere, elias:wheres-the-bear}.

With that said, we note two limitations associated with our evaluation methodology. First, both of the datasets we chose are relatively small in size, and do gain some advantage from MongoDB's internal caching, as previously stated. Second, while our workloads are widely used in the machine learning community, they are not benchmarks in and of themselves, and our results may not be directly comparable to other work using these workloads. As such, further investigations into this technique may wish to use a peer-reviewed machine learning benchmark, such as Deep500 or DDLBench~\cite{DEEP500,DDLBench}.

While \sysname presents a promising way to perform distributed training with centralized bucket storage, there are a number of potential directions one can explore to further understand and improve the performance of cloud-based data loading.
First, some of the performance savings were realized due to MongoDB's internal caching, which stores frequently accessed samples in memory \cite{mongodb:wiredtiger-caching-docs}. It would be useful to quantify what portion of our savings comes from this mechanism. Second, though \sysname is agnostic of architecture, measuring how it performs in other architectures, such as a parameter server architecture \cite{parameter_server_li}, would underscore its usefulness in other applications. Last, cost could be further optimized by cutting down on the number of necessary bucket API calls. Namely, Class A requests could be cut down by caching the list of items in a bucket locally, and Class B requests could be cut down by grouping samples into \textit{super-samples}: collections of multiple samples. These super-samples would reduce the number of fetches needed to download a node's partition from the bucket, but the partitioning strategy would need to be altered to account for them.

\section{Related Work}
\label{chapter:related-work}

    There has been limited research into optimizing the data-loading aspect of distributed training. The most relevant work we found to our research was by Yang et al. \cite{data_transfer_yang}. This paper suggests a novel approach to caching, in which every node keeps a local cache of mini-batches, and uses this cache to determine which samples it should fetch from the storage server. If a node's cache does not contain a balanced number of samples from different mini-batches, it may exchange samples with another node. We initially attempted to adapt this paper's methodology to the cloud; however, Yang et al.\@'s paper does not take into account the limitations of a cloud environment, including low bandwidth between nodes. Unlike their work, though, \sysname does not need to concern itself with the load capacity of the storage medium, as GCP's buckets auto-scale~\cite{website:bucket-scaling}.

    In addition, some existing research relating to serverless computing demonstrated some of the ways that our approach could be applied more generally. The Cirrus framework \cite{Cirrus}, for instance, implements both caching and pre-fetching from buckets in their approach, and demonstrated significant gains when pre-fetching was added. They also use other techniques that we considered during our design process, such as batching multiple samples in the object store. Cirrus demonstrates that these approaches can reduce training time in serverless environments, and complements our study of saving time on data loading, even outside a serverless framework.

    Wang et al. propose an interesting caching idea that could serve as a good middle-layer on top of \sysname. Their system, InfiniCache, creates a cache by storing samples in-memory in serverless functions, and then allows them to become idle. These functions stay warm and ready to serve requests for several hours, despite their idle state, which allows the system to operate as a cheap and efficient cache \cite{infinicache}. Though on its own InfiniCache serves a different purpose than \sysname, it may serve as a good intermediary between \sysname and the storage buckets, due to its low cost and high performance of retrieval.

    Some other work has been done to accelerate data augmentation~\cite{augmentation_zolnouri}, which is the transformation process that data must undergo before being fed into the neural network. This is an important step of the data loading process, and was not something that \sysname attempts to accelerate. Its integration could help further improve data loading time.

\section{Conclusion}
\label{chapter:conclusion}

In this paper, we investigated the use of classical techniques, caching and pre-fetching, to support distributed training with cloud-based storage. We implemented a prototype of our system, \sysname, on top of the PyTorch Framework and evaluated its effectiveness on Google Cloud Platform. While other work has focused on speeding up training or coordination between nodes, we aim to fill the research gap in speeding up data loading, specifically within cloud-based environments that use bucket storage. As Section \ref{chapter:results} details, our system can perform as well as storing data locally on each node under the right conditions, such as when training a more complicated model that takes longer per epoch. While increasing the cache space per node (and the corresponding fetch size/pre-fetch threshold) would also speed up data loading, this research chooses to consider the storage space per node as a constraint and to characterize how caching and pre-fetching can be used effectively within this given amount of storage.

With that in mind, we want to reiterate that the goal of our research is not to find better configurations for virtual machines, but rather how to improve the data-loading within those configurations. In doing this, we hope that this approach will easily apply to smaller-scale distributed deep learning operations. While adding more nodes, increasing storage, and renting faster hardware will almost always help reduce epoch run time, they are beyond the scope of this research. This also helps our research remain applicable to scenarios where storing large amounts of data on disk may not be possible, such as serverless workloads.

We measured the data loading time of our model under different conditions and compare these measurements to baselines where pieces of our model were selectively removed. This comparison allows us to decompose the observed benefits into what time savings come from caching, pre-fetching, and setting a pre-fetch threshold. Our evaluation demonstrates that \sysname is effective, and that each of its three major components are necessary to fully realize its potential to speed up data loading in bucket-based distributed deep learning.

\section*{Acknowledgment}
We thank all anonymous reviewers and shepherds for their comments and suggestions which helped improve this paper.
This work is supported in part by National Science Foundation grants CNS-\#1755659 and CNS-\#1815619, WPI CS Department funding, and Google Cloud Platform free credits.

\bibliographystyle{IEEEtran}
\bibliography{proposal}

\end{document}